\documentclass[aps,prl,reprint,superscriptaddress,amsmath,amssymb,floatfix]{revtex4-2}

\usepackage{setspace}
\usepackage{graphicx}
\usepackage{bm}
\usepackage[utf8]{inputenc}
\usepackage[colorlinks=true,allcolors=blue]{hyperref}
\usepackage[T1]{fontenc}

\begin{document}

\title{Bound-State Beta Decay of $\mathbf{\mathrm{^{205}{Tl}^{81+}}}$ Ions and the LOREX Project}

\author{	R.~S.~Sidhu}
\email{ragan.sidhu@ed.ac.uk}
\affiliation{School  of  Physics  and  Astronomy, The University  of  Edinburgh,  EH9 3FD  Edinburgh, United Kingdom}
\affiliation{GSI Helmholtzzentrum f\"{u}r Schwerionenforschung, Planckstra{\ss}e 1, 64291 Darmstadt, Germany}
\affiliation{Max-Planck-Institut f\"{u}r Kernphysik, 69117 Heidelberg, Germany}

\author{	G.~Leckenby}
\affiliation{TRIUMF, Vancouver, British Columbia V6T 2A3, Canada}
\affiliation{Department of Physics and Astronomy, University of British Columbia, Vancouver, BC V6T 1Z1, Canada}

\author{	R.~J.~Chen}
\email{r.chen@gsi.de}
\affiliation{GSI Helmholtzzentrum f\"{u}r Schwerionenforschung, Planckstra{\ss}e 1, 64291 Darmstadt, Germany}
\affiliation{Max-Planck-Institut f\"{u}r Kernphysik, 69117 Heidelberg, Germany}
\affiliation{Institute of Modern Physics, Chinese Academy of Sciences, 730000 Lanzhou, People's Republic of China}

\author{	R.~Mancino}
\email{riccardo.mancino@matfyz.cuni.cz}
\altaffiliation[Present address: ]{Faculty of Mathematics and Physics, Charles University, Prague, Czech Republic}
\affiliation{Institut f\"{u}r Kernphysik (Theoriezentrum), Fachbereich
  Physik, Technische  Universit\"{a}t Darmstadt, Schlossgartenstra{\ss}e 2, 64289 Darmstadt, Germany}
\affiliation{GSI Helmholtzzentrum f\"{u}r Schwerionenforschung, Planckstra{\ss}e 1, 64291 Darmstadt, Germany}

\author{	T.~Neff }
\affiliation{GSI Helmholtzzentrum f\"{u}r Schwerionenforschung, Planckstra{\ss}e 1, 64291 Darmstadt, Germany}

\author{	Yu.~A.~Litvinov	}
\affiliation{GSI Helmholtzzentrum f\"{u}r Schwerionenforschung, Planckstra{\ss}e 1, 64291 Darmstadt, Germany}
\affiliation{Helmholtz Forschungsakademie Hessen f\"ur FAIR (HFHF), GSI
    Helmholtzzentrum f\"ur Schwerionenforschung,
  Planckstra{\ss}e~1,
    64291 Darmstadt, Germany}

\author{	G.~Mart{\'i}nez-Pinedo	}
\affiliation{GSI Helmholtzzentrum f\"{u}r Schwerionenforschung, Planckstra{\ss}e 1, 64291 Darmstadt, Germany}
\affiliation{Institut f\"{u}r Kernphysik (Theoriezentrum), Fachbereich
  Physik, Technische  Universit\"{a}t Darmstadt, Schlossgartenstra{\ss}e 2, 64289 Darmstadt, Germany}
\affiliation{Helmholtz Forschungsakademie Hessen f\"ur FAIR (HFHF), GSI
    Helmholtzzentrum f\"ur Schwerionenforschung,
  Planckstra{\ss}e~1,
    64291 Darmstadt, Germany}

\author{	G.~Amthauer}
\affiliation{Department of Chemistry and Physics of Materials, University of Salzburg, Jakob-Haringer-Strasse 2a, 5020 Salzburg, Austria}

\author{	M.~Bai	}
\affiliation{GSI Helmholtzzentrum f\"{u}r Schwerionenforschung, Planckstra{\ss}e 1, 64291 Darmstadt, Germany}

\author{	K.~Blaum	}
\affiliation{Max-Planck-Institut f\"{u}r Kernphysik, 69117 Heidelberg, Germany}

\author{	B.~Boev}
\affiliation{University of {\v S}tip, Faculty of Mining and Geology, Goce Delčev 89, 92000 {\v S}tip, North Macedonia}

\author{	F.~Bosch}
\thanks{Deceased}
\affiliation{GSI Helmholtzzentrum f\"{u}r Schwerionenforschung, Planckstra{\ss}e 1, 64291 Darmstadt, Germany}

\author{	C.~Brandau	}
\affiliation{GSI Helmholtzzentrum f\"{u}r Schwerionenforschung, Planckstra{\ss}e 1, 64291 Darmstadt, Germany}

\author{	V.~Cvetkovi{\'c}}
\affiliation{University of Belgrade, Faculty of Mining and Geology, \DJ ušina 7, 11000 Belgrade, Serbia}

\author{	T.~Dickel	}
\affiliation{GSI Helmholtzzentrum f\"{u}r Schwerionenforschung, Planckstra{\ss}e 1, 64291 Darmstadt, Germany}
\affiliation{II. Physikalisches Institut, Justus-Liebig-Universit\"{a}t Gie{\ss}en, 35392 Gie{\ss}en, Germany}

\author{	I.~Dillmann	}
\affiliation{TRIUMF, Vancouver, British Columbia V6T 2A3, Canada}
\affiliation{Department of Physics and Astronomy, University of Victoria, Victoria, British Columbia V8P 5C2, Canada}

\author{	D.~Dmytriiev	}
\affiliation{GSI Helmholtzzentrum f\"{u}r Schwerionenforschung, Planckstra{\ss}e 1, 64291 Darmstadt, Germany}

\author{	T.~Faestermann	}
\affiliation{Physik Department, Technische Universität M\"{u}nchen, D-85748 Garching, Germany}

\author{	O.~Forstner	}
\affiliation{GSI Helmholtzzentrum f\"{u}r Schwerionenforschung, Planckstra{\ss}e 1, 64291 Darmstadt, Germany}

\author{	B.~Franczak	}
\affiliation{GSI Helmholtzzentrum f\"{u}r Schwerionenforschung, Planckstra{\ss}e 1, 64291 Darmstadt, Germany}

\author{	H.~Geissel	}
\thanks{Deceased}
\affiliation{GSI Helmholtzzentrum f\"{u}r Schwerionenforschung, Planckstra{\ss}e 1, 64291 Darmstadt, Germany}
\affiliation{II. Physikalisches Institut, Justus-Liebig-Universit\"{a}t Gie{\ss}en, 35392 Gie{\ss}en, Germany}

\author{	R.~Gernhäuser}
\affiliation{Physik Department, Technische Universität M\"{u}nchen, D-85748 Garching, Germany}

\author{	J.~Glorius	}
\affiliation{GSI Helmholtzzentrum f\"{u}r Schwerionenforschung, Planckstra{\ss}e 1, 64291 Darmstadt, Germany}

\author{	C.~J.~Griffin	}
\affiliation{TRIUMF, Vancouver, British Columbia V6T 2A3, Canada}

\author{	A.~Gumberidze	}
\affiliation{GSI Helmholtzzentrum f\"{u}r Schwerionenforschung, Planckstra{\ss}e 1, 64291 Darmstadt, Germany}

\author{	E.~Haettner	}
\affiliation{GSI Helmholtzzentrum f\"{u}r Schwerionenforschung, Planckstra{\ss}e 1, 64291 Darmstadt, Germany}

\author{	P.-M.~Hillenbrand	}
\affiliation{GSI Helmholtzzentrum f\"{u}r Schwerionenforschung, Planckstra{\ss}e 1, 64291 Darmstadt, Germany}
\affiliation{I. Physikalisches Institut, Justus-Liebig-Universit\"{a}t Gie{\ss}en, 35392 Gie{\ss}en, Germany}

\author{	P.~Kienle	}
\thanks{Deceased}
\affiliation{Physik Department, Technische Universität M\"{u}nchen, D-85748 Garching, Germany}

\author{	W.~Korten}
\affiliation{IRFU, CEA, Universit\'{e} Paris-Saclay, Gif-sur-Yvette, 91191, France}

\author{	Ch.~Kozhuharov}
\affiliation{GSI Helmholtzzentrum f\"{u}r Schwerionenforschung, Planckstra{\ss}e 1, 64291 Darmstadt, Germany}

\author{	N.~Kuzminchuk	}
\affiliation{GSI Helmholtzzentrum f\"{u}r Schwerionenforschung, Planckstra{\ss}e 1, 64291 Darmstadt, Germany}

\author{	K.~Langanke}
\affiliation{GSI Helmholtzzentrum f\"{u}r Schwerionenforschung, Planckstra{\ss}e 1, 64291 Darmstadt, Germany}

\author{	S.~Litvinov	}
\affiliation{GSI Helmholtzzentrum f\"{u}r Schwerionenforschung, Planckstra{\ss}e 1, 64291 Darmstadt, Germany}

\author{	E.~Menz	}
\affiliation{GSI Helmholtzzentrum f\"{u}r Schwerionenforschung, Planckstra{\ss}e 1, 64291 Darmstadt, Germany}

\author{	T.~Morgenroth	}
\affiliation{GSI Helmholtzzentrum f\"{u}r Schwerionenforschung, Planckstra{\ss}e 1, 64291 Darmstadt, Germany}

\author{	C.~Nociforo	}
\affiliation{GSI Helmholtzzentrum f\"{u}r Schwerionenforschung, Planckstra{\ss}e 1, 64291 Darmstadt, Germany}

\author{	F.~Nolden	}
\thanks{Deceased}
\affiliation{GSI Helmholtzzentrum f\"{u}r Schwerionenforschung, Planckstra{\ss}e 1, 64291 Darmstadt, Germany}

\author{	M.~K.~Pavićević}
\affiliation{Department of Chemistry and Physics of Materials, University of Salzburg, Jakob-Haringer-Strasse 2a, 5020 Salzburg, Austria}

\author{	N.~Petridis	}
\affiliation{GSI Helmholtzzentrum f\"{u}r Schwerionenforschung, Planckstra{\ss}e 1, 64291 Darmstadt, Germany}

\author{	U.~Popp	}
\affiliation{GSI Helmholtzzentrum f\"{u}r Schwerionenforschung, Planckstra{\ss}e 1, 64291 Darmstadt, Germany}

\author{	S.~Purushothaman	}
\affiliation{GSI Helmholtzzentrum f\"{u}r Schwerionenforschung, Planckstra{\ss}e 1, 64291 Darmstadt, Germany}

\author{	R.~Reifarth	}
\affiliation{J.W. Goethe Universit\"{a}t, 60438 Frankfurt, Germany}

\author{	M.~S.~Sanjari	}
\affiliation{GSI Helmholtzzentrum f\"{u}r Schwerionenforschung, Planckstra{\ss}e 1, 64291 Darmstadt, Germany}

\author{	C.~Scheidenberger	}
\affiliation{GSI Helmholtzzentrum f\"{u}r Schwerionenforschung, Planckstra{\ss}e 1, 64291 Darmstadt, Germany}
\affiliation{II. Physikalisches Institut, Justus-Liebig-Universit\"{a}t Gie{\ss}en, 35392 Gie{\ss}en, Germany}
\affiliation{Helmholtz Research Academy Hesse for FAIR (HFHF), GSI Helmholtz Center for Heavy Ion Research, Campus Gießen, 35392 Gießen, Germany}

\author{	U.~Spillmann	}
\affiliation{GSI Helmholtzzentrum f\"{u}r Schwerionenforschung, Planckstra{\ss}e 1, 64291 Darmstadt, Germany}

\author{	M.~Steck	}
\affiliation{GSI Helmholtzzentrum f\"{u}r Schwerionenforschung, Planckstra{\ss}e 1, 64291 Darmstadt, Germany}

\author{	Th.~St\"{o}hlker	}
\affiliation{GSI Helmholtzzentrum f\"{u}r Schwerionenforschung, Planckstra{\ss}e 1, 64291 Darmstadt, Germany}

\author{	Y.~K.~Tanaka	}
\affiliation{High Energy Nuclear Physics Laboratory, RIKEN, 2-1 Hirosawa, Wako, Saitama 351-0198, Japan}

\author{	M.~Trassinelli}
\affiliation{Institut des NanoSciences de Paris,
CNRS, Sorbonne Université, Paris, France}

\author{	S.~Trotsenko	}
\affiliation{GSI Helmholtzzentrum f\"{u}r Schwerionenforschung, Planckstra{\ss}e 1, 64291 Darmstadt, Germany}

\author{	L.~Varga	}
\affiliation{Physik Department, Technische Universität M\"{u}nchen, D-85748 Garching, Germany}
\affiliation{GSI Helmholtzzentrum f\"{u}r Schwerionenforschung, Planckstra{\ss}e 1, 64291 Darmstadt, Germany}

\author{	M.~Wang	}
\affiliation{Institute of Modern Physics, Chinese Academy of Sciences, 730000 Lanzhou, People's Republic of China}

\author{	H.~Weick	}
\affiliation{GSI Helmholtzzentrum f\"{u}r Schwerionenforschung, Planckstra{\ss}e 1, 64291 Darmstadt, Germany}

\author{	P.~J.~Woods	}
\affiliation{School  of  Physics  and  Astronomy, The University  of Edinburgh,  EH9 3FD  Edinburgh, United Kingdom}

\author{	T.~Yamaguchi}
\affiliation{Saitama University, Saitama 338-8570, Japan}

\author{	Y.~H.~Zhang	}
\affiliation{Institute of Modern Physics, Chinese Academy of Sciences, 730000 Lanzhou, People's Republic of China}

\author{	J.~Zhao } 
\affiliation{GSI Helmholtzzentrum f\"{u}r Schwerionenforschung, Planckstra{\ss}e 1, 64291 Darmstadt, Germany}

\author{	K.~Zuber } 
\affiliation{Institut f{\"u}r Kern- und Teilchenphysik, Technische Universit{\"a}t Dresden, Zellescher Weg 19, 01062 Dresden, Germany}

\collaboration{E121 and LOREX Collaborations}

\date{\today}

\begin{abstract}
  Stable $^{205}$Tl ions have the lowest known energy threshold for capturing electron neutrinos ($\nu_e$) of ${ E}_{\nu_e}\ge50.6$\,keV. 
  The Lorandite Experiment (LOREX), proposed in the 1980s, aims at obtaining the longtime averaged solar neutrino flux by utilizing natural deposits of Tl-bearing lorandite ores.
  To determine the $\nu_e$ capture cross section, it is required to know the strength of the weak transition connecting the ground state of $^{205}$Tl and the 2.3 keV first excited state in $^{205}$Pb. 
  The only way to experimentally address this transition is to measure the bound-state beta decay ($\beta_{b}$) of fully ionized $\mathrm{^{205}Tl^{81+}}$ ions.
  After three decades of meticulous preparation, the half-life of the $\beta_{b}$ decay of $\mathrm{^{205}Tl^{81+}}$ has been measured to be $291_{-27}^{+33}$ days using the Experimental Storage Ring (ESR) at GSI, Darmstadt. 
  The longer measured half-life compared to theoretical estimates reduces the expected signal-to-noise ratio in the LOREX, thus challenging its feasibility.
 \end{abstract}


\maketitle

\textit{Introduction}---In the past 4.6 billion years~\cite{bonanno2002age}, the Sun has produced
an abundant amount of energy through nuclear fusion
reactions~\cite{RevModPhys.83.195}. 
As a result, a huge amount of electron neutrinos ($\nu_e$) have been, and continue to be, emitted
from the Sun's
interior~\cite{Bahcall_1988,Haxton.Hamish.Serenelli:2013}. 
The main
contribution ($\sim$~91$\%$) comes from the proton-proton ($pp$)
fusion reaction. 
Other
contributions include neutrinos from weak decays of $\mathrm{^7Be}$ ($\sim$~7$\%$), $\mathrm{^8B}$ ($\sim$~0.02$\%$), and CNO cycle isotopes (see Table~\ref{table1}).
The neutrino spectroscopy provides important information
on the structure of our Sun and serves as a critical means to
test solar
models. 

Accounting for most of the solar neutrinos, $pp$ neutrinos are
extremely challenging to detect on Earth owing to their low energy (0~$\leq$ $E_{{\nu_{e}}}$~$\leq$~422~keV). 
The first neutrino experiments~\cite{RevModPhys.71.1213, abdurashitov1994results, anselmann1992solar} observed neutrino yields much lower than expected by solar models, which was termed as the solar-neutrino problem. 
In 1976,~\citet{freedman1976solar} studied all weak transitions and identified the $^{205}$Tl-$^{205}$Pb pair to have the lowest threshold energy for $\nu_e$ capture of just 50.6\,keV~\cite{Wang_2021}. 
They proposed to search for natural deposits of lorandite ores (TlAs$\mathrm{S_2}$) containing $^{205}$Tl (natural abundance of 70.5$\%$) and measure the amount of created $^{205}$Pb, thereby determining the solar neutrino flux, $\Phi_{\nu_e}$, averaged over the age of the ore.

The
Lorandite Experiment (LOREX)~\cite{pavicevic1988lorandite} was conceived by Pavi\'{c}evi\'{c} in 1983, which relied on the rich lorandite deposits discovered in the mine of Allchar (North Macedonia)~\cite{pavicevic2010new,pavicevic2018lorandite}. Although the solar-neutrino problem was solved through the discovery
of neutrino oscillations~\cite{pontecorvo1968neutrino} at the Sudbury Neutrino
Observatory (SNO)~\cite{Ahmad.Allen.ea:2002} and the Super-Kamiokande (Super-K)~\cite{suzuki2019super}, and the present $\Phi_{\nu_e}$ value was obtained by the BOREXINO experiment~\cite{borexino2014neutrinos,borexino2018comprehensive}, 
the possibility to obtain a longtime averaged $\Phi_{\nu_e}$ remains appealing, which justified the continuation of LOREX~\cite{emmi_lorex}.

Of the various geochemical solar-neutrino experiments, such as
$^{81}$Br-$^{81}$Kr~\cite{kuzminov1988new}, $^{98}$Mo-$^{98}$Tc~\cite{bahcall1988will}, and others~\cite{hahn2008radiochemical}, 
LOREX stands as the final geochemical neutrino project~\cite{Kirsten-book}. 
LOREX combines the geological extraction and characterization of lorandite ore, its chemical purification, and an accelerator-based measurement of $^{205}$Pb concentration to determine the $\nu_e$ capture cross section of $^{205}$Tl.
In this Letter, we report the first direct measurement of the bound-state beta decay ($\beta_{b}$) of fully ionized $\mathrm{^{205}Tl^{81+}}$, which is indispensable for the determination of the $\nu_e$ capture cross section. 
This decay rate is also important for the astrophysical slow neutron-capture process (the \textit{s} process), which is discussed elsewhere~\cite{Leckenby_sprocess}.

The capture of solar neutrinos transforms the stable
$\mathrm{^{205}Tl}$ atom (ground state, $I^{\pi}=1/2^{+}$) via the ($\nu_e,e^-$) reaction into the
radionuclide $\mathrm{^{205}Pb}$, where the first excited state
[$E^*$\,=\,2.3~keV~\cite{nndc}, $I^\pi$\,=\,$1/2^-$, $T_{1/2}=24.2(4)$\,$\mu$s] is predominantly populated due
to the nuclear selection rules. This state promptly decays via internal conversion to the ground state of $\mathrm{^{205}Pb}$ [$I^{\pi}=5/2^{-}$, $T_{1/2}$\,=\,17.0(9) million years (Myr)~\cite{kondev2021nubase2020} ], making the measurement of its weak decay unfeasible. 
The only direct method to determine the nuclear matrix element of this weak transition is currently the measurement of the bound-state beta decay of fully ionized $\mathrm{^{205}{Tl}^{81+}}$ to $\mathrm{^{205}{Pb}^{81+}}$~\cite{e121proposal}. 

In $\beta_b$ decay~\cite{daudel1947possibilite,bahcall1961theory}, a neutron transforms into a proton, and an electron is created in a bound atomic state instead of being emitted to a continuum, and thus, a monochromatic electron antineutrino is produced in a free state.

\textit{Experiment and analysis}---The measurement 
was performed at GSI Helmholtzzentrum f\"{u}r Schwerionenforschung in Darmstadt, which is presently the only facility where such measurements are possible. 
Since $^{205}$Tl atoms are stable, a straightforward approach would be to produce the $^{205}$Tl beam directly from the ion source as was done in the first $\beta_b$-decay measurements on $\mathrm{^{163}Dy}$~\cite{jung1992first} and $^{187}$Re\,\cite{bosch1996observation}. 
However, Tl vapors are poisonous.
Therefore, $^{205}$Tl$^{81+}$ ions had to be produced in a nuclear reaction.
An enriched $\mathrm{^{206}Pb}$ material was utilized.
The $\mathrm{^{206}Pb^{67+}}$ projectile beams were accelerated to an energy of 11.4~MeV/u in the linear accelerator UNILAC and then injected into the heavy-ion synchrotron  SIS-18,  where they were further accelerated up to an energy of 678~MeV/u.  
The $\mathrm{^{206}Pb}$ beams were extracted with an average intensity of ${10}^9$~particles per pulse and impinged on a 1607~mg/$\mathrm{{cm}^2}$ thick $\mathrm{^9{Be}}$ production target backed with 223~mg/$\mathrm{{cm}^2}$ niobium located at the entrance of the Fragment Separator (FRS)~\cite{geissel1992gsi}. 
$\mathrm{^{205}Tl^{81+}}$ ions were produced via a single proton knockout reaction along with numerous other reaction products. The cocktail fragment beam was analyzed by the FRS.
The critical contaminant fragment is the isobaric hydrogenlike (H-like) $^{205}$Pb$^{81+}$ ions, which are the same as the daughter ions of the $\beta_b$ decay of interest. 
The fully ionized $\mathrm{^{205}Tl^{81+}}$ ions were centered throughout the FRS.
An aluminum degrader with a thickness of 735~mg/$\mathrm{{cm}^2}$ was used in the middle focal plane of the FRS.
Owing to the $Z^2$ dependence of the energy losses and subsequent momentum analysis, the distributions of $^{205}$Pb$^{81+}$ and $\mathrm{^{205}Tl^{81+}}$ were nearly completely spatially separated at the exit of the FRS.

About $10^4$ $\mathrm{^{205}Tl^{81+}}$ ions per spill were transmitted through the FRS and injected into the Experimental Storage Ring (ESR)~\cite{franzke1987esr} at 400~MeV/u. 
The proportion of contaminant $\mathrm{^{205}Pb^{81+}}$ ions amounted to 0.1$\%$ of the $\mathrm{^{205}Tl^{81+}}$ intensity. 
The adopted production and separation scheme is analogous to the one employed in the measurement of $\beta_b$ decay of $\mathrm{^{207}Tl^{81+}}$~\cite{ohtsubo2005simultaneous}. However, the larger dispersion in the second half of the FRS used in this experiment enabled us to reduce the contamination by about an order of magnitude.

According to theoretical predictions, the half-life of $^{205}$Tl$^{81+}$ could have been as long as 1~yr~\cite{takahashi_nuclear_1983}.
Therefore, to achieve higher statistics, accumulation of $\mathrm{^{205}Tl^{81+}}$ ions was necessary. 
The storage acceptance of the ESR is $\Delta B\rho / B\rho$ $\sim$ $\pm1.5$\%.
Fresh fragment beams were injected on an outer orbit of the ESR, where they were stochastically precooled~\cite{stochastic_cooling}. 
The cooled beam was then shifted by applying a radio-frequency pulse to an inner orbit of the ESR forming a stack.  
This procedure was repeated up to several hundred cycles, where each newly injected beam was added to the stack. The beam was continuously cooled by an electron cooler~\cite{electron_cooling, steck2004electron} (electron current $I_e=200$~mA).
Once the desired intensity was achieved [$N_{\rm Tl}(0)\approx1$--$2 \times 10^6$], stochastic cooling was switched off and the stacked beam was shifted from the inner orbit to the central orbit by ramping the dipole magnets. 
To allow $^{205}$Tl$^{81+}$ to decay, we let the beam circulate in the ultrahigh vacuum of the ESR ($\approx 10^{-10}$\,mbar) for different storage times, ranging from 0 to 10 hr, while the current of the electron cooler was reduced to $I_e=20$~mA to minimize the recombination rate with electrons. 

\begin{figure}[!t]
    \centering
    \includegraphics[width=\linewidth]{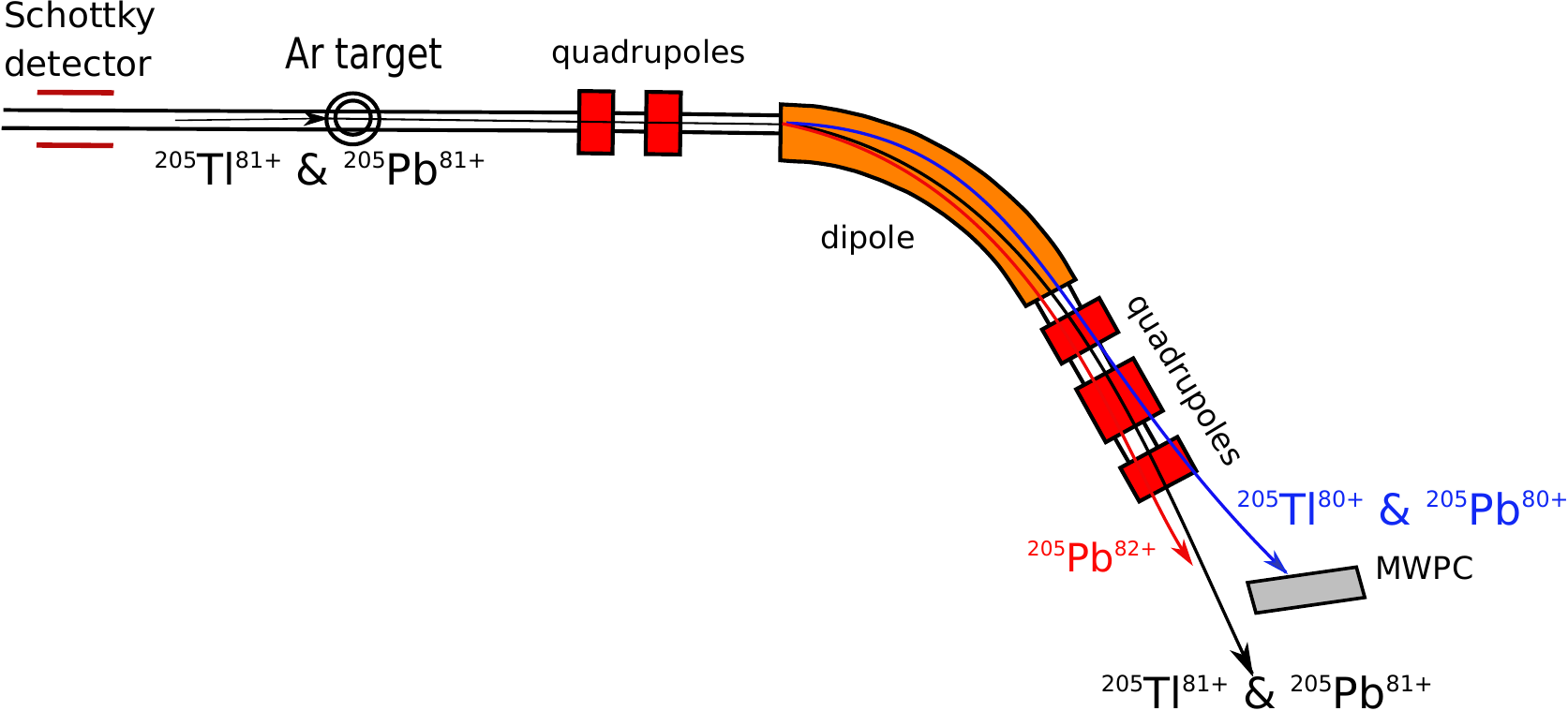}
    {\caption{Schematic of the experimental setup at the ESR from the gas jet target to the next dipole magnet. The $\beta_b$-decayed $\mathrm{^{205}Pb^{81+}}$ daughter ions were stripped to the $q=82+$ state, where they followed a different orbit to the main beam within the acceptance of the ring. Ions with $q= 80+$ were deflected and counted by a MWPC~\cite{klepper2003particle}. Both, the $\mathrm{^{205}Tl^{81+}}$ and $\mathrm{^{205}Pb^{82+}}$ ions were counted by a non destructive 245 MHz resonant Schottky detector~\cite{nolden2011schottky,sanjari2013resonant,Sanjari2013}.} 
    \label{figure-1}}
\end{figure}

\begin{figure}[!t]
\centering
\includegraphics[width=\linewidth]{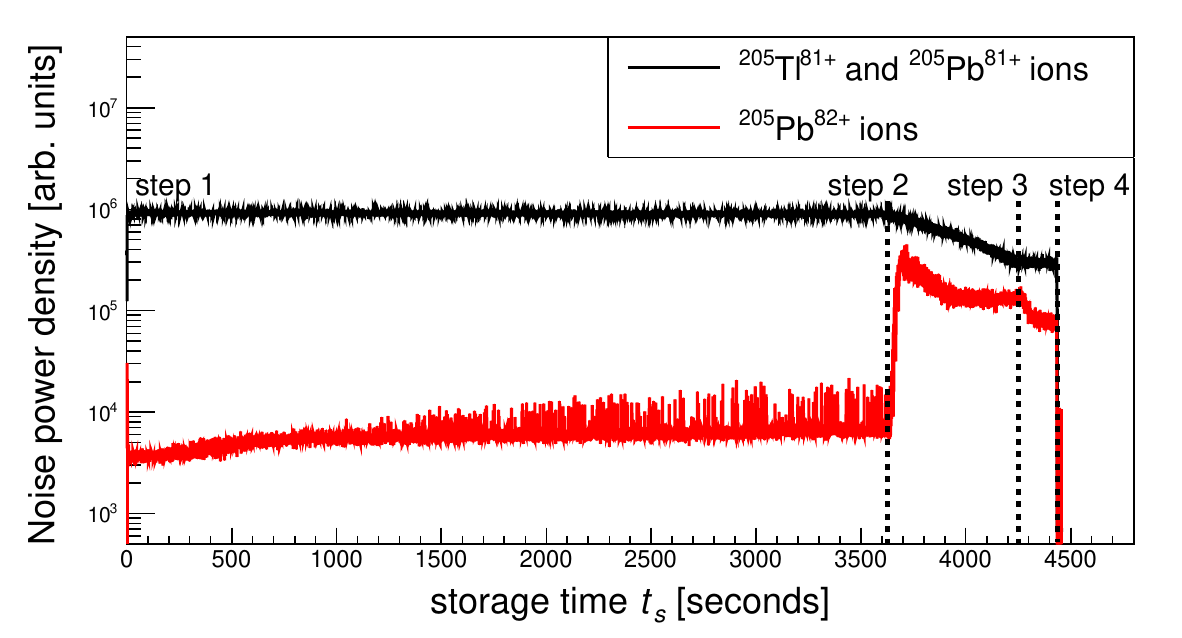}
{\caption{Intensities of $q=81+$ and $q=82+$ ions during an example 1 hr storage measurement. The storage time (step 1 $\rightarrow$ step 2), gas jet operation (step 2 $\rightarrow$ step 3), cooling (step 3 $\rightarrow$ step 4), and detection (step 4).}
\label{figure-2}}
\end{figure}

During the storage time, some of the $\mathrm{^{205}Tl^{81+}}$ ions decayed by $\beta_b$ decay to H-like $\mathrm{^{205}Pb^{81+}}$ ions with the electron created in the \textit{K} shell.
Because of a small $Q_{\beta_b}$(\textit{K}) value of only 31.1(5)~keV~\cite{Wang_2021,kondev2021nubase2020,kramida2023nist}, the two beams were indistinguishable by detectors in the ESR.
To reveal the $\beta_b$-daughter ions at the end of each storage time, an argon (Ar) gas jet target~\cite{petridis2015prototype,Petridis2014} with a density $\sim$ (2--$4) \times 10^{12}$~atoms/$\mathrm{{cm}^2}$ was switched on for a time interval of 10\,min.
As a result, the electron from $\mathrm{^{205}Pb^{81+}}$ ions was stripped off, transforming H-like $\mathrm{^{205}Pb^{81+}}$ ions to the fully ionized $\mathrm{^{205}Pb^{82+}}$ ions. 
Thereby, the mass to charge ($m/q$) ratio was altered and $\mathrm{^{205}Pb^{82+}}$ ions jumped to an inner orbit of the ring (as shown in Fig.~\ref{figure-1}). 
After the application of the gas target, the ions were electron cooled for 100\,s and then counted for a further 40~s, see Fig.~\ref{figure-2}. 

The number of $\beta_b$-daughter ions $N_{\mathrm{Pb}}$($t_{s}$) grows in proportion to the storage time $t_{s}$ provided that $t_{s}$ is small with respect to the half-life. 
For that case, $N_{\mathrm{Pb}}$($t_{s}$) is given to a good approximation by the relation~\cite{jung1992first,sidhu2021thesis}
\begin{multline}
 \frac{N_{\rm Pb} (t_{s})}{N_{\rm Tl}(t_{s})}   =  
\frac{\lambda_{b}}{\gamma}  t_{s}
[1 +  \frac{1}{2} (\lambda_{\rm Tl}^{cc} 
-\lambda_{\rm Pb}^{cc} ) t_{s}+ \cdot \cdot \cdot] \\ +  \frac{N_{\rm Pb} (0)}{N_{\rm Tl}(0)} e^{(\lambda_{\rm Tl}^{cc}  -\lambda_{\rm Pb}^{cc} ) t_{s}} , 
\label{eq1}
\end{multline}
where $N_{\rm Tl}(t_{s})$  and $N_{\rm Pb}(t_{s})$ are, respectively, the numbers of $\mathrm{^{205}Tl^{81+}}$ and $\mathrm{^{205}Pb^{81+}}$ ions at time, $t_{s}$, the end of the storage, with further details on their determination provided in~\cite{Sidhu_supplement}. 
The ratio $N_{\rm Pb} (0)/N_{\rm Tl}(0)$ is the initial contamination.
The detection and identification of stored ions was done with a non destructive 245~MHz Schottky resonator~\cite{nolden2011schottky,sanjari2013resonant,Sanjari2013}, where the Fourier transformed noise from the detector reveals revolution frequencies and intensities of all individual ionic species\,\cite{Litvinov-2011, Steck-2020}.
The Lorentz factor was $\gamma=1.429(1)$.
The rates $\lambda_{\rm Tl}^{cc}$ and  $\lambda_{\rm Pb}^{cc}$ account for storage losses of
$^{205}\rm Tl^{81+}$  and $^{205}\rm Pb^{81+}$  ions, respectively, due to the atomic charge changing ($cc$)
processes in the ESR.
$\lambda_{\rm Tl}^{cc}$ = $4.34(6) \times 10^{-5}$~$\rm s^{-1}$ was measured with $\mathrm{^{205}Tl^{81+}}$ beam. 
Theoretical rates of radiative recombination (RR)~\cite{PhysRevA.45.7894} were used to determine the difference 
$\lambda_{\rm Tl}^{cc}  -\lambda_{\rm Pb}^{cc}=\lambda_{\rm Tl}^{cc}(1-\rm RR_{\rm Pb^{81+}}/ \rm RR_{\rm Tl^{81+}})=3.47(5)\mathrm{_{stat}}(87)\mathrm{_{syst}} \times 10^{-6}~\mathrm{s}^{-1}$. 

H-like $\mathrm{^{205}Pb^{81+}}$ ions may capture an electron from the gas jet atoms and leave the acceptance of the ESR. 
Thus, the number of $\mathrm{^{205}Pb^{82+}}$ ions detected on the inner orbit needs to be corrected by the ratio $(\sigma_{\rm I,\rm Pb} + \sigma_{\rm C, \rm Pb})/\sigma_{\rm I, \rm Pb}$~\cite{jung1992first}, 
where $\sigma_{\rm I,\rm Pb}$  and $\sigma_{\rm C,\rm Pb}$ are the ionization and capture cross sections, respectively.
The ratio $(\sigma_{\rm I,\rm Pb} + \sigma_{\rm C, \rm Pb})/\sigma_{\rm I, \rm Pb}=1.425(3)\mathrm{_{stat}}(13)\mathrm{_{syst}}$ was obtained with a $^{206}\rm Pb^{81+}$ beam~\cite{sidhu2021thesis}. $\mathrm{^{206}Pb^{80+}}$ ions that captured an electron in the gas jet target were counted by a multiwire proportional chamber (MWPC)~\cite{klepper2003particle} placed outside the ring vacuum behind a 25~$\mu$m stainless foil (see Fig.\,\ref{figure-1}), while 
the total number of ions was monitored by a direct current current transformer~\cite{reeg2001current}. 

\begin{figure}[!t]
\centering
\includegraphics[width=\linewidth]{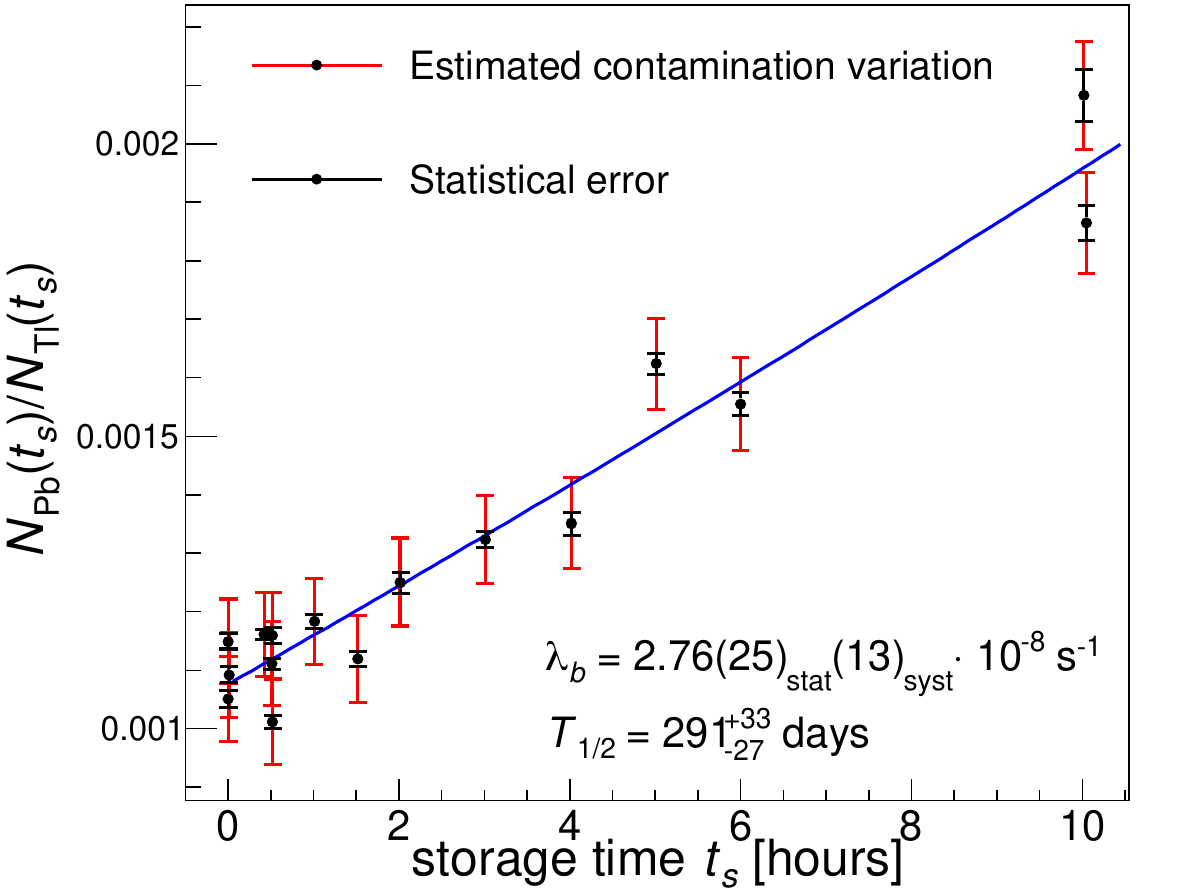}
\caption{$\mathrm{^{205}Pb^{81+}}$  to  $\mathrm{^{205}Tl^{81+}}$  ratio  as  a  function  of  storage  time $t_{s}$. Equation~(\ref{eq1}) is used to fit (blue line) and determine the decay constant for the bound-state beta decay $\beta_b$. The error bars in black represent the statistical errors, whereas the error bars in red represent the estimated contamination variation, which accounts for the amount of $\mathrm{^{205}Pb^{81+}}$ ions injected in the ESR from the FRS.}
\label{figure-3}     
\end{figure}
 
A Monte Carlo error propagation method was used to consistently handle the applied corrections. 
In particular, the variation in the initial contamination of $\mathrm{^{205}Pb^{81+}}$ ions was estimated from the data itself. 
To estimate the contamination variation, the $\chi^2$ distribution
was sampled for each Monte Carlo run, and then a value for the variation
was determined~\cite{Sidhu_supplement}.
Subsequently, the dataset was fitted to Eq.~(\ref{eq1}), see Fig.~\ref{figure-3}, and the median $\beta_b$-decay rate in the ion rest frame was determined to be $\lambda_{b}$ = 2.76(25)$\mathrm{_{stat}}$(13)$\mathrm{_{syst}} \times 10^{-8}$~$\rm s^{-1}$, which corresponds to a half-life of $T_{1/2}$ =  $291_{-27}^{+33}$~days. 
More details on the data analysis can be found in~\cite{sidhu2021thesis,leckenby2023analysis}.

\textit{From bound decay to neutrino capture}---Because of the conservation of energy, the possible final end states of the $\beta_{b}$ decay of $\mathbf{\mathrm{^{205}{Tl}}}$ can only be the ground state $\mathrm{^{205}Pb}$ ($I^{\pi}=5/2^{-})$ and the first excited state $\mathrm{^{205}Pb}$ ($I^{\pi}=1/2^{-})$. 
Hence, the total $\beta_b$-decay rate can be expressed as $\lambda_b$ = $\lambda_b^{5/2^-}$ $+$ $\lambda_b^{1/2^-}$. 
From the available rate of the inverse process, i.e., the decay via electron capture of $\mathrm{^{205}Pb}$ into $\mathbf{\mathrm{^{205}{Tl}}}$, it is possible to determine that $\lambda_b^{5/2^-} = 1.44(8)\times 10^{-13}$ $\rm s^{-1}$, and thus it has a negligible contribution. 

Furthermore, considering that only the decay with the creation of an
electron in the $K$ shell of $\mathrm{^{205}Pb}$ is energetically
allowed, it is then possible to express the total decay rate as
$\lambda_b \simeq \lambda_b^{1/2^-} = \ln(2) f_K
C_K/\mathcal{K}$, where
$\mathcal{K} = 2\overline{\mathcal{F}t} = 6144.5 (37)$~s~ is the
corrected $\beta$-decay constant that includes the independent
radiative correction $\Delta^V_R$ as detailed
in~\cite{hardy_superallowed_2020},
$f_K=\pi Q_{\beta_b}^2 \beta_K^2/(2 m_e^2 c^4)$ is the phase space
for $\beta_b$ decay,
$m_e$ is the electron mass, and
$\beta_K$ is the Coulomb amplitude of the $K$-shell electron wave
function~\cite{bambynek_orbital_1977}. 
Using $\beta_K^2 =
5.567$ for H-like
$^{205}$Pb computed with the flexible atomic code~\cite{FAC-code}, and $Q_{\beta_b}=31.1(5)$~keV, we
obtain $f_K=
0.032(1)$, which together with the measured decay rate gives a value
for the nuclear shape factor for $\beta_b$ decay of $C_K= 7.6(8)\times10^{-3}$, corresponding to $\log ft = \log (\mathcal{K}/C_K) = 5.91(5)$. 
This shape factor has also been used as a basis for the
calculations of the stellar weak rates involving
$^{205}$Pb and $^{205}$Tl in~\cite{Leckenby_sprocess}. 
A previous theoretical study found $C_K = 0.010$~\cite{Warburton_1991}.  

\begin{figure}
\centering
\includegraphics[width=\linewidth]{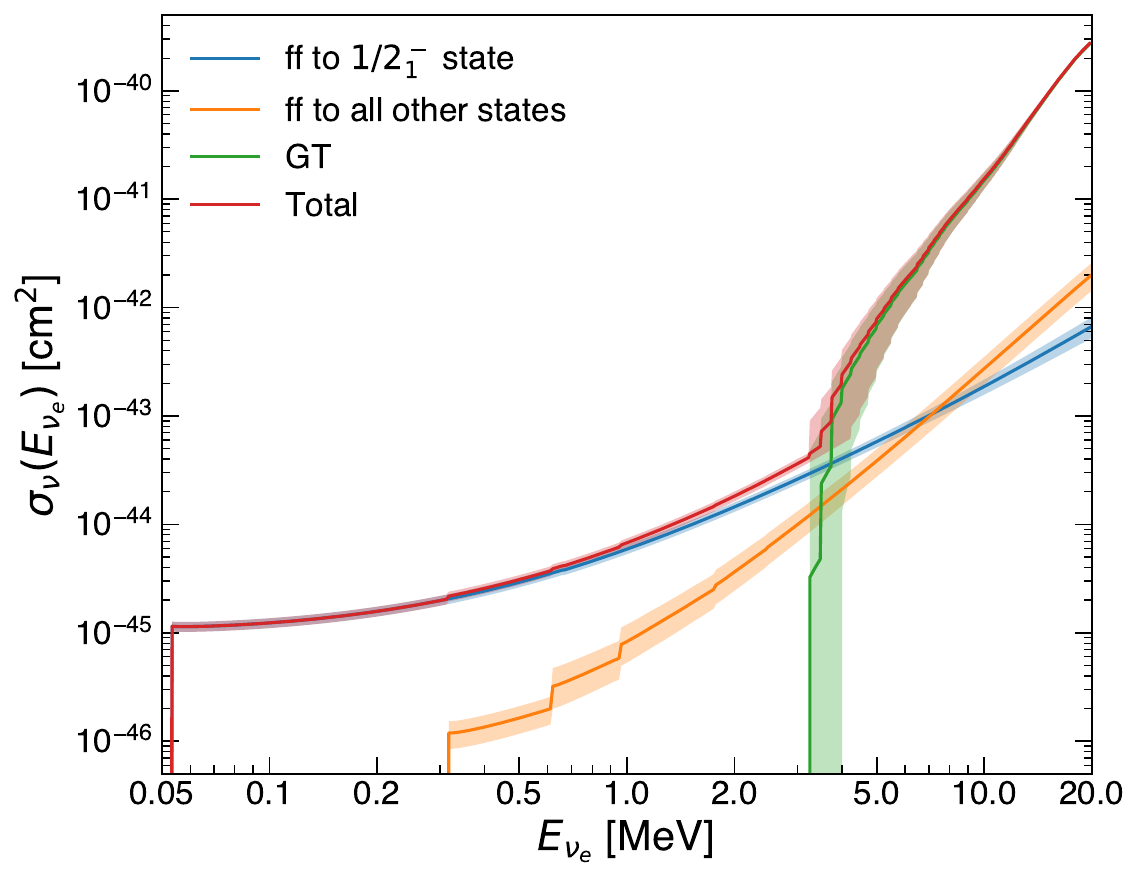}
\caption{Contributions to the total cross section for neutrino capture $\sigma(E_{\nu_e})$ from the $^{205}$Tl ground state to $^{205}$Pb, along with their relative uncertainties. Note that the total cross section is mainly determined by experimental information for all the relevant energies.\label{sigma_nu}}
\end{figure}

$C_K$ contains all relevant nuclear structure information that can be
expressed in terms of the usual nuclear matrix elements for
first-forbidden (ff) decays using the Behrens and Bühring formalism
\cite{behrens1982electron}. 
However, while the emitted
neutrino is monoenergetic in $\beta_b$ decay, in the case of neutrino
capture, the equivalent nuclear shape factor,
$C_{1/2^-_1}(E_{\nu_e})$ for the first excited state in $\mathrm{^{205}Pb}$ ($I^{\pi}=1/2^{-}$), must be evaluated for all values of the captured neutrino
energy,
$E_{\nu_e}$, as given by the solar-neutrino spectrum. This requires us to
disentangle the contributions of the different nuclear matrix elements (see~\cite{Sidhu_supplement} for further details). 
For this purpose,
we adopted the shell-model framework and used the \texttt{NATHAN}
code~\cite{Caurier_2005} together with the Kuo-Herling hole-hole
interaction~\cite{Warburton_1991}.
Based on the shell-model calculations, we include first-forbidden contributions to the
solar-neutrino capture to excited states in
$^{205}$Pb, beyond the ground state and the first excited state. 
As seen in Fig.~\ref{sigma_nu}, we find that first-forbidden transitions are sufficient to describe the cross section up to neutrino energies around 3~MeV. 
For higher neutrino energies, positive parity states in $^{205}$Pb become accessible, and it is necessary to consider allowed Gamow-Teller (GT) transitions. 
We estimated this contribution based on charge-exchange data from the $^{205}$Tl$(p,n){}^{205}$Pb reaction~\cite{Krofcheck_1987}. 

Figure~\ref{sigma_nu} shows the energy dependent cross section for neutrino capture $\sigma(E_{\nu_e})$, decomposed into the main contributions. 
The contribution to the cross section by the transition to $^{205}$Pb$(1/2^-_1)$ dominates at low neutrino energies ($E_{\nu_e} < 3$ MeV). 
The cross section has been integrated over the different solar-neutrino fluxes to obtain the different contributions to the solar-neutrino capture rate, as given in Table~\ref{table1} (see~\cite{Sidhu_supplement}). We account for neutrino oscillations using the solar-neutrino survival factors extracted from Fig.~14.3 in~\cite{Navas_2024}.  
We also include the estimates from
the work of Bahcall and Ulrich~\cite{Bahcall_1988}. The latter have
been rescaled to the values of neutrino fluxes used in this work. 
In parenthesis, we provide the contribution to the neutrino capture of the transition to $^{205}$Pb$(1/2^-_1)$. 
It represents the dominant contribution for all neutrino sources except for $^8$B and $hep$ neutrinos, whose rates are mainly determined by Gamow-Teller transitions. 

We obtain a total neutrino capture rate of $92\pm10\pm10$ solar neutrino unit (SNU), where the first error accounts for nuclear uncertainties and the second for the uncertainties in the neutrino survival factors. We notice that the capture cross section for
$pp$ and $^7$Be neutrinos is significantly lower than the one of Bahcall and Ulrich. Lacking a measurement for the shape factor,
Bahcall and Ulrich assumed a $\log ft = 5.7$ value for the first-forbidden transition to $^{205}$Pb($1/2^-_1$) from systematics. Our
total neutrino capture rate is significantly higher than the one reported in~\cite{Kostensalo_2019} ($62.2 \pm 8.6$ SNU).
This is most likely due to their neglect of Gamow-Teller
contributions and an underestimation of the $pp$-neutrino rate.
Further details on the individual contributions of the different neutrino fluxes can be found in~\cite{Sidhu_supplement}.

\begin{center}
\begin{table}
  \caption{Contributions of individual neutrino fluxes to the solar-neutrino capture rate on $^{205}$Tl expressed in SNU. Neutrino
    oscillations are considered. The quoted errors account only for nuclear uncertainties. In the work of Bahcall and
    Ulrich~\cite{Bahcall_1988}, the capture of neutrinos from the
    $^{13}$N, $^{15}$O, and $pep$ fluxes are considered in the total
    sum but not provided individually. In
parentheses, the contribution to the neutrino capture of the
transition to $^{205}$Pb$(1/2^-_1)$ is given.\label{table1}}
  \begin{ruledtabular}
    \renewcommand{\arraystretch}{1.2}
\begin{tabular}{ccc}
Flux & Present Letter & Ref.~\cite{Bahcall_1988} \\
 \hline
${pp}$ & $ 64 \pm 7$ $\, (63 \pm 7)$ & $98$ \\
${^7\text{Be}}$ & $14 \pm 2$ $\, (12 \pm 1)$ & $19$ \\
${^{13}\text{N}}$  & $0.85  \pm 0.11$ $\, (0.74 \pm 0.05)$ & - \\
${^{15}\text{O}}$  & $0.95  \pm 0.11$ $\, (080 \pm 0.11)$ & - \\
${^{17}\text{F}}$  & $0.021  \pm 0.002$ $\, (0.018  \pm 0.002)$ & - \\
${pep}$  & $0.69  \pm 0.09$ $\, (0.56 \pm 0.04)$ & - \\
${^8\text{B}}$  & $11\pm 2 $ $\, (0.18 \pm 0.02)$ & 13 \\
${hep}$  & $0.26 ^{+0.06} _{-0.03}$ $\, (0.0019 \pm 0.0003)$ & - \\
\hline
${\text{Total}}$ & $92\pm 10$ $\, (77 \pm 8)$ & $135$ \\
\end{tabular}
\end{ruledtabular}
\end{table}
\end{center}
\textit{LOREX implications.}---The goal of the LOREX project is to
determine the longtime average of the solar neutrino
flux $\Phi_{\nu}$ via the neutrino capture reaction. 
The average
neutrino flux over the exposure time \textit{a}, which represents the
age of lorandite since its mineralization, is estimated using the
activation equation~\cite{pavicevic2010new}: $\Phi_{\nu}$ =
$N^{-1}(T-B)(\sigma \epsilon$)$^{-1} \lambda$[$1 - \mathrm{exp}(-\lambda a$)]$^{-1}$,
where \textit{N} is the total number of $^{205}$Tl atoms, \textit{T}
is the total number of $^{205}$Pb atoms, \textit{B} is the
background-induced number of $^{205}$Pb atoms (mainly produced by
cosmic muons and natural radioactivity~\cite{pavicevic2010new,pavicevic2012status}), $\sigma$ denotes the neutrino capture cross section,
$\epsilon$ represents the overall detection efficiency, and $\lambda$
is the decay constant of $^{205}$Pb.  
Considering the total neutrino
capture rate of $92\pm10\pm10$~SNU determined here, 
the geological age of
\textit{a} = 4.31(2) Myr~\cite{neubauer200940ar}, the electron capture probability
$\lambda$ of $^{205}$Pb back to $^{205}$Tl as $\lambda$ = $4.01(16)\times 10^{-8}$~yr$^{-1}$~\cite{kondev2021nubase2020}, the molar mass \textit{M} of
lorandite = 343 g/mol~\cite{pavicevic2018lorandite}, and the number of $^{205}$Tl atoms to be $1.25\times 10^{21}(2.5 \times 10^{19}$) atoms/g of lorandite~\cite{pavicevic2012status}, we expect the time-integrated number of solar-neutrino induced $^{205}$Pb atoms ($T-B$) to be 14(4) atoms/g
of lorandite. 
This obtained value agrees within 1$\sigma$ with the value quoted in
\cite{pavicevic2018lorandite}, i.e., 22(7) $^{205}$Pb atoms/g of lorandite, where a
theoretical value of 146~SNU was used for the neutrino capture rate.

Presently, a total of 405.5 g of pure lorandite crystals, with a purity
$>99$\%, have been extracted from 10.5 tons of raw ore body of Crven Dol Allchar deposit. 
These lorandite samples are stored in a shallow underground Felsenkeller laboratory in
Dresden to ensure minimal production of background-induced $^{205}$Pb
atoms \cite{sidhu2021thesis}. 
The next significant challenge for the LOREX project is to extract the total number
of Pb atoms (\textit{T}) from the collected lorandite minerals. 
For this purpose, dedicated
accelerator-based experiments are being discussed
\cite{e121proposal,pavicevic2010new,pavicevic2012status}.
Additionally, it is crucial to subtract the background-induced $^{205}$Pb atoms (\textit{B}) which depend on factors
such as the erosion rate and paleo depth of the lorandite site from where the samples are collected and can be calculated as a function of the depth
of the lorandite location~\cite{pavicevic2012status}. 

\textit{Conclusions}---In conclusion, the measurement of the half-life of the bound-state beta decay of fully ionized $^{205}$Tl$^{81+}$ ions has been successfully accomplished. 
This experiment was one of the main motivations for the construction of the SIS-FRS-ESR facilities at GSI~\cite{kienle1988esr}. 
After 30 years since its proposal, continuous advancements in accelerator technologies, beam manipulation techniques, and detector performances have finally led to the realization of this measurement.
The measured $\beta_b$-decay half-life is significantly longer, $291_{-27}^{+33}$~days, than previous theoretical estimates based on systematics: 122 days~\cite{Taka1987} and 52.43 days~\cite{Liu}. 
By using the new estimated rate, the solar-neutrino capture cross section has been calculated, resulting in a correspondingly lower value than previously anticipated.
This has severe consequences on the geochemical neutrino experiment LOREX, which aims to determine the averaged solar neutrino flux over the last 4.31(2) Myr. 
Taking the estimated concentration of 14(4) solar neutrino induced $^{205}$Pb atoms/g, a total of about 5677(1622) $^{205}$Pb atoms are expected in the entire available lorandite sample of about 405.5\,g, which corresponds to a signal-to-background ratio of only 3.5$\sigma$ (assuming detection efficiency of 100\%). 
Overall, this makes any statistically significant determination of $^{205}$Pb concentration due to solar neutrinos highly unlikely. The sensitivity might be boosted by collecting significantly larger lorandite samples, preferably at higher paleo depths to reduce cosmic and geology related systematic uncertainties~\cite{pavicevic2010new,pavicevic2018lorandite,pavicevic2012status}.

\textit{Acknowledgments}---The results presented here are based on the experiment G-17-E121, which was performed at the FRS-ESR facilities at the GSI Helmholtzzentrum f{\"u}r Schwerionenforschung, Darmstadt (Germany) in the framework of FAIR Phase-0.
Fruitful discussions and support from 
U.~Battino, D.~Bemmerer, S.~Cristallo, B.~S.~Gao, R.~Grisenti, S.~Hagmann, W.~F.~Henning, A.~Karakas, T.~Kaur, O.~Klepper, W.~Kutschera, M.~Lestinsky, M.~Lugaro, B.~Meyer, A.~Ozawa, V.~Pejovi{\'c}, M.~Pignatari, 
D.~Schneider, T.~Suzuki, B.~Szányi, K.~Takahashi, S.~Yu.~Torilov, X.~L.~Tu, D.~Vescovi, P.~M.~Walker, E. Wiedner, N.~Winckler, A.~Yagüe Lopéz, and X.~H.~Zhou are greatly acknowledged. 
The authors thank the GSI accelerator
team for providing excellent technical support, in particular, to R. Heß, C. Peschke, and J. Roßbach from the GSI beam cooling group. 
We thank the ExtreMe Matter Institute EMMI at GSI, Darmstadt, for support in the framework of an EMMI Rapid Reaction Task Force meeting.
This work was
supported by the European Research Council (ERC) under the EU’s
Horizon 2020 research and innovation program (Grant Agreement Nos.
682841 “ASTRUm" and 654002 “ENSAR2"); the Natural Sciences and Engineering Research Council of Canada (NSERC) (NSERC Discovery Grant
No. SAPIN-2019-00030); the Excellence Cluster ORIGINS from the
German Research Foundation DFG (Excellence Strategy
EXC-2094–390783311); the Science and Technology Facilities
Council (STFC) (Grant No. ST/P004008/1); and the Sumitomo Foundation, Mitsubishi Foundation, JSPS
KAKENHI Nos. 26287036, 17H01123, 23KK0055. 
R.M. and G.M.P. acknowledge
support by the Deutsche Forschungsgemeinschaft (DFG, German Research Foundation)---Project-ID 279384907 - SFB 1245.
Y.H.Z. acknowledges the support from EMMI at GSI in the form of an EMMI Visiting Professorship. R.M. acknowledges the support provided by Grant No. 23-06439S of the Czech Science Foundation.

The authors declare
that they have no known competing financial interests or
personal relationships that could have appeared to influence
the work reported in this Letter.


%

\end{document}



\title{Supplemental Material for: Bound-State Beta Decay of ${}^{205}$Tl${}^{81+}$ Ions and the LOREX Project}



\author{	R.~S.~Sidhu}
\email{ragan.sidhu@ed.ac.uk}
\affiliation{School  of  Physics  and  Astronomy, The University  of  Edinburgh,  EH9 3FD  Edinburgh, United Kingdom}
\affiliation{GSI Helmholtzzentrum f\"{u}r Schwerionenforschung, Planckstra{\ss}e 1, 64291 Darmstadt, Germany}
\affiliation{Max-Planck-Institut f\"{u}r Kernphysik, 69117 Heidelberg, Germany}

\author{	G.~Leckenby}
\affiliation{TRIUMF, Vancouver, British Columbia V6T 2A3, Canada}
\affiliation{Department of Physics and Astronomy, University of British Columbia, Vancouver, BC V6T 1Z1, Canada}

\author{	R.~J.~Chen}
\email{r.chen@gsi.de}
\affiliation{GSI Helmholtzzentrum f\"{u}r Schwerionenforschung, Planckstra{\ss}e 1, 64291 Darmstadt, Germany}
\affiliation{Max-Planck-Institut f\"{u}r Kernphysik, 69117 Heidelberg, Germany}
\affiliation{Institute of Modern Physics, Chinese Academy of Sciences, 730000 Lanzhou, People's Republic of China}

\author{	R.~Mancino}
\email{riccardo.mancino@matfyz.cuni.cz}
\altaffiliation[Present address: ]{Faculty of Mathematics and Physics, Charles University, Prague, Czech Republic}
\affiliation{Institut f\"{u}r Kernphysik (Theoriezentrum), Fachbereich
  Physik, Technische  Universit\"{a}t Darmstadt, Schlossgartenstra{\ss}e 2, 64289 Darmstadt, Germany}
\affiliation{GSI Helmholtzzentrum f\"{u}r Schwerionenforschung, Planckstra{\ss}e 1, 64291 Darmstadt, Germany}

\author{	T.~Neff }
\affiliation{GSI Helmholtzzentrum f\"{u}r Schwerionenforschung, Planckstra{\ss}e 1, 64291 Darmstadt, Germany}

\author{	Yu.~A.~Litvinov	}
\affiliation{GSI Helmholtzzentrum f\"{u}r Schwerionenforschung, Planckstra{\ss}e 1, 64291 Darmstadt, Germany}
\affiliation{Helmholtz Forschungsakademie Hessen f\"ur FAIR (HFHF), GSI
    Helmholtzzentrum f\"ur Schwerionenforschung,
  Planckstra{\ss}e~1,
    64291 Darmstadt, Germany}

\author{	G.~Mart{\'i}nez-Pinedo	}
\affiliation{GSI Helmholtzzentrum f\"{u}r Schwerionenforschung, Planckstra{\ss}e 1, 64291 Darmstadt, Germany}
\affiliation{Institut f\"{u}r Kernphysik (Theoriezentrum), Fachbereich
  Physik, Technische  Universit\"{a}t Darmstadt, Schlossgartenstra{\ss}e 2, 64289 Darmstadt, Germany}
\affiliation{Helmholtz Forschungsakademie Hessen f\"ur FAIR (HFHF), GSI
    Helmholtzzentrum f\"ur Schwerionenforschung,
  Planckstra{\ss}e~1,
    64291 Darmstadt, Germany}

\author{	G.~Amthauer}
\affiliation{Department of Chemistry and Physics of Materials, University of Salzburg, Jakob-Haringer-Strasse 2a, 5020 Salzburg, Austria}

\author{	M.~Bai	}
\affiliation{GSI Helmholtzzentrum f\"{u}r Schwerionenforschung, Planckstra{\ss}e 1, 64291 Darmstadt, Germany}

\author{	K.~Blaum	}
\affiliation{Max-Planck-Institut f\"{u}r Kernphysik, 69117 Heidelberg, Germany}

\author{	B.~Boev}
\affiliation{University of {\v S}tip, Faculty of Mining and Geology, Goce Delčev 89, 92000 {\v S}tip, North Macedonia}

\author{	F.~Bosch}
\thanks{Deceased}
\affiliation{GSI Helmholtzzentrum f\"{u}r Schwerionenforschung, Planckstra{\ss}e 1, 64291 Darmstadt, Germany}

\author{	C.~Brandau	}
\affiliation{GSI Helmholtzzentrum f\"{u}r Schwerionenforschung, Planckstra{\ss}e 1, 64291 Darmstadt, Germany}

\author{	V.~Cvetkovi{\'c}}
\affiliation{University of Belgrade, Faculty of Mining and Geology, \DJ ušina 7, 11000 Belgrade, Serbia}

\author{	T.~Dickel	}
\affiliation{GSI Helmholtzzentrum f\"{u}r Schwerionenforschung, Planckstra{\ss}e 1, 64291 Darmstadt, Germany}
\affiliation{II. Physikalisches Institut, Justus-Liebig-Universit\"{a}t Gie{\ss}en, 35392 Gie{\ss}en, Germany}

\author{	I.~Dillmann	}
\affiliation{TRIUMF, Vancouver, British Columbia V6T 2A3, Canada}
\affiliation{Department of Physics and Astronomy, University of Victoria, Victoria, British Columbia V8P 5C2, Canada}

\author{	D.~Dmytriiev	}
\affiliation{GSI Helmholtzzentrum f\"{u}r Schwerionenforschung, Planckstra{\ss}e 1, 64291 Darmstadt, Germany}

\author{	T.~Faestermann	}
\affiliation{Physik Department, Technische Universität M\"{u}nchen, D-85748 Garching, Germany}

\author{	O.~Forstner	}
\affiliation{GSI Helmholtzzentrum f\"{u}r Schwerionenforschung, Planckstra{\ss}e 1, 64291 Darmstadt, Germany}

\author{	B.~Franczak	}
\affiliation{GSI Helmholtzzentrum f\"{u}r Schwerionenforschung, Planckstra{\ss}e 1, 64291 Darmstadt, Germany}

\author{	H.~Geissel	}
\thanks{Deceased}
\affiliation{GSI Helmholtzzentrum f\"{u}r Schwerionenforschung, Planckstra{\ss}e 1, 64291 Darmstadt, Germany}
\affiliation{II. Physikalisches Institut, Justus-Liebig-Universit\"{a}t Gie{\ss}en, 35392 Gie{\ss}en, Germany}

\author{	R.~Gernhäuser}
\affiliation{Physik Department, Technische Universität M\"{u}nchen, D-85748 Garching, Germany}

\author{	J.~Glorius	}
\affiliation{GSI Helmholtzzentrum f\"{u}r Schwerionenforschung, Planckstra{\ss}e 1, 64291 Darmstadt, Germany}

\author{	C.~J.~Griffin	}
\affiliation{TRIUMF, Vancouver, British Columbia V6T 2A3, Canada}

\author{	A.~Gumberidze	}
\affiliation{GSI Helmholtzzentrum f\"{u}r Schwerionenforschung, Planckstra{\ss}e 1, 64291 Darmstadt, Germany}

\author{	E.~Haettner	}
\affiliation{GSI Helmholtzzentrum f\"{u}r Schwerionenforschung, Planckstra{\ss}e 1, 64291 Darmstadt, Germany}

\author{	P.-M.~Hillenbrand	}
\affiliation{GSI Helmholtzzentrum f\"{u}r Schwerionenforschung, Planckstra{\ss}e 1, 64291 Darmstadt, Germany}
\affiliation{I. Physikalisches Institut, Justus-Liebig-Universit\"{a}t Gie{\ss}en, 35392 Gie{\ss}en, Germany}

\author{	P.~Kienle	}
\thanks{Deceased}
\affiliation{Physik Department, Technische Universität M\"{u}nchen, D-85748 Garching, Germany}

\author{	W.~Korten}
\affiliation{IRFU, CEA, Universit\'{e} Paris-Saclay, Gif-sur-Yvette, 91191, France}

\author{	Ch.~Kozhuharov}
\affiliation{GSI Helmholtzzentrum f\"{u}r Schwerionenforschung, Planckstra{\ss}e 1, 64291 Darmstadt, Germany}

\author{	N.~Kuzminchuk	}
\affiliation{GSI Helmholtzzentrum f\"{u}r Schwerionenforschung, Planckstra{\ss}e 1, 64291 Darmstadt, Germany}

\author{	K.~Langanke}
\affiliation{GSI Helmholtzzentrum f\"{u}r Schwerionenforschung, Planckstra{\ss}e 1, 64291 Darmstadt, Germany}

\author{	S.~Litvinov	}
\affiliation{GSI Helmholtzzentrum f\"{u}r Schwerionenforschung, Planckstra{\ss}e 1, 64291 Darmstadt, Germany}

\author{	E.~Menz	}
\affiliation{GSI Helmholtzzentrum f\"{u}r Schwerionenforschung, Planckstra{\ss}e 1, 64291 Darmstadt, Germany}

\author{	T.~Morgenroth	}
\affiliation{GSI Helmholtzzentrum f\"{u}r Schwerionenforschung, Planckstra{\ss}e 1, 64291 Darmstadt, Germany}

\author{	C.~Nociforo	}
\affiliation{GSI Helmholtzzentrum f\"{u}r Schwerionenforschung, Planckstra{\ss}e 1, 64291 Darmstadt, Germany}

\author{	F.~Nolden	}
\thanks{Deceased}
\affiliation{GSI Helmholtzzentrum f\"{u}r Schwerionenforschung, Planckstra{\ss}e 1, 64291 Darmstadt, Germany}

\author{	M.~K.~Pavićević}
\affiliation{Department of Chemistry and Physics of Materials, University of Salzburg, Jakob-Haringer-Strasse 2a, 5020 Salzburg, Austria}

\author{	N.~Petridis	}
\affiliation{GSI Helmholtzzentrum f\"{u}r Schwerionenforschung, Planckstra{\ss}e 1, 64291 Darmstadt, Germany}

\author{	U.~Popp	}
\affiliation{GSI Helmholtzzentrum f\"{u}r Schwerionenforschung, Planckstra{\ss}e 1, 64291 Darmstadt, Germany}

\author{	S.~Purushothaman	}
\affiliation{GSI Helmholtzzentrum f\"{u}r Schwerionenforschung, Planckstra{\ss}e 1, 64291 Darmstadt, Germany}

\author{	R.~Reifarth	}
\affiliation{J.W. Goethe Universit\"{a}t, 60438 Frankfurt, Germany}

\author{	M.~S.~Sanjari	}
\affiliation{GSI Helmholtzzentrum f\"{u}r Schwerionenforschung, Planckstra{\ss}e 1, 64291 Darmstadt, Germany}

\author{	C.~Scheidenberger	}
\affiliation{GSI Helmholtzzentrum f\"{u}r Schwerionenforschung, Planckstra{\ss}e 1, 64291 Darmstadt, Germany}
\affiliation{II. Physikalisches Institut, Justus-Liebig-Universit\"{a}t Gie{\ss}en, 35392 Gie{\ss}en, Germany}
\affiliation{Helmholtz Research Academy Hesse for FAIR (HFHF), GSI Helmholtz Center for Heavy Ion Research, Campus Gießen, 35392 Gießen, Germany}

\author{	U.~Spillmann	}
\affiliation{GSI Helmholtzzentrum f\"{u}r Schwerionenforschung, Planckstra{\ss}e 1, 64291 Darmstadt, Germany}

\author{	M.~Steck	}
\affiliation{GSI Helmholtzzentrum f\"{u}r Schwerionenforschung, Planckstra{\ss}e 1, 64291 Darmstadt, Germany}

\author{	Th.~St\"{o}hlker	}
\affiliation{GSI Helmholtzzentrum f\"{u}r Schwerionenforschung, Planckstra{\ss}e 1, 64291 Darmstadt, Germany}

\author{	Y.~K.~Tanaka	}
\affiliation{High Energy Nuclear Physics Laboratory, RIKEN, 2-1 Hirosawa, Wako, Saitama 351-0198, Japan}

\author{	M.~Trassinelli}
\affiliation{Institut des NanoSciences de Paris,
CNRS, Sorbonne Université, Paris, France}

\author{	S.~Trotsenko	}
\affiliation{GSI Helmholtzzentrum f\"{u}r Schwerionenforschung, Planckstra{\ss}e 1, 64291 Darmstadt, Germany}

\author{	L.~Varga	}
\affiliation{Physik Department, Technische Universität M\"{u}nchen, D-85748 Garching, Germany}
\affiliation{GSI Helmholtzzentrum f\"{u}r Schwerionenforschung, Planckstra{\ss}e 1, 64291 Darmstadt, Germany}

\author{	M.~Wang	}
\affiliation{Institute of Modern Physics, Chinese Academy of Sciences, 730000 Lanzhou, People's Republic of China}

\author{	H.~Weick	}
\affiliation{GSI Helmholtzzentrum f\"{u}r Schwerionenforschung, Planckstra{\ss}e 1, 64291 Darmstadt, Germany}

\author{	P.~J.~Woods	}
\affiliation{School  of  Physics  and  Astronomy, The University  of Edinburgh,  EH9 3FD  Edinburgh, United Kingdom}

\author{	T.~Yamaguchi}
\affiliation{Saitama University, Saitama 338-8570, Japan}

\author{	Y.~H.~Zhang	}
\affiliation{Institute of Modern Physics, Chinese Academy of Sciences, 730000 Lanzhou, People's Republic of China}

\author{	J.~Zhao } 
\affiliation{GSI Helmholtzzentrum f\"{u}r Schwerionenforschung, Planckstra{\ss}e 1, 64291 Darmstadt, Germany}

\author{	K.~Zuber } 
\affiliation{Institut f{\"u}r Kern- und Teilchenphysik, Technische Universit{\"a}t Dresden, Zellescher Weg 19, 01062 Dresden, Germany}

\collaboration{E121 and LOREX Collaborations}


\date{\today}

\pacs{}

\maketitle


\textit{Determination of ion numbers}---To calculate the $\beta_b$ decay constant $\lambda_b$ in Eq.~(1) of the main paper, the number of mother $\mathrm{^{205}Tl^{81+}}$ ($N_{\rm Tl}(t_{s})$) and $\beta_b$-daughter $\mathrm{^{205}Pb^{81+}}$ ($N_{\rm Pb}(t_{s})$) ions needs to be measured at the end of the storage measurement, $t_s$.
These ion numbers were determined at the end of the storage measurement, 
step 2 in Fig.~2 in the main paper, from the measured ion numbers at step 4 using the following equations 
\begin{equation}
N_{\rm Tl}(t_{s}) = N_{\rm Tl}^{\rm step 4}(t_{s})   \mathrm{S_C}(t_s)   \frac{1}{ e^{-\lambda_{\rm Tl} \Delta t} },  \label{eq2}
\end{equation}
\begin{equation}
N_{\rm Pb}(t_{s}) = N_{\rm Pb}^{\rm step 4}(t_{s}) \frac{\sigma_{\rm I,\rm Pb}+\sigma_{\rm C, \rm Pb}}{\sigma_{\rm I, \rm Pb}} \frac{1}{ \mathrm{R_C}}   \frac{1}{1- e^{-\lambda_{\rm Pb} \Delta t}}, 
\label{eq3}
\end{equation}
where $\Delta t$ = 10 min is the time duration for which the gas jet target was turned on, and $\sigma_{\rm I,\rm Pb}$  and $\sigma_{\rm C,\rm Pb}$ are the ionization and capture cross sections for $^{205}$Pb, respectively.
The decay constants $\lambda_{\rm Tl}$ and $\lambda_{\rm Pb}$ were measured individually for each storage period to account for fluctuations in target density.
$\rm{S_C}$($t_s$) accounts for a saturation effect, which was directly proportional to the beam intensity, caused by a mismatched amplifier switch in the data acquisition system\,\cite{trageser2015ntcap,christian_thesis}.
It has been calibrated with a low-intensity measurement and a current transformer, 
more details of which can be found in \cite{sidhu2021thesis,leckenby2023analysis}. 
$\rm{R_C}$ is the correction coefficient used 
to account for the resonance response of the 245 MHz Schottky detector. 
The $\rm{R_C}$ was calibrated with multiple ion species, the frequencies of which were modified when moving the beam after the accumulation was completed~\cite{sidhu2021thesis,leckenby2023analysis}.

\vspace{1ex}
\textit{Estimated contamination variation}---The $^{205}\mathrm{Pb}^{81+}$ ions created by one neutron knockout in the projectile fragmentation reaction in the FRS provide a significant source of contamination for the measurement of the $\beta_b$ decay of $^{205}\mathrm{Tl}^{81+}$. 
From Fig.~3 in the main paper, the $t_s=0$ intercept indicates that the contamination was at the $\approx0.11\%$ level, which was larger than the $\beta_b$ signal even after 10~hrs of storage. 
In previous $\beta_b$-decay experiments~\cite{jung1992first,bosch1996observation}, the authors were able to purge the $\beta_b$-daughter contaminants by turning on the gas jet target before storage, which reduced their contamination to essentially zero.
After stacking reached saturation, they achieved up to $10^8$ stored ions. As they were working with primary beams, losing parent ions during the purging procedure was not an issue. 
Whereas, we achieved $\sim$~1-$2\times10^6$ ions as we were stacking secondary beams and could not afford the loss in beam intensity.

The contaminant $^{205}\mathrm{Pb}^{81+}$ ions from the FRS were expected to be stable across storage times and thus provided an additive offset to the observed signal that we fit as a free parameter. 
However, after controlling for all other sources of error, the $\chi^2$ of our data was 302.7, which is well beyond the 95\% confidence interval of [6.6,23.7] for 14 degrees of freedom. 
Thus, in the absence of any other plausible explanation, we concluded that the contamination from the FRS must have varied stochastically on the 1\% level. 
This was supported by estimates of the magnet stability throughout the FRS chain. 
It was challenging to get an accurate estimate of the size of the variation though, because $99.9\%$ of contaminant $^{205}\mathrm{Pb}^{81+}$ was blocked by the FRS slits, so the magnet variation would be affecting the tails of the straggling distribution at $>3\sigma$. 
At these extremes, the tails are most likely non-Gaussian and could behave in non-linear ways.

\begin{figure}
\centering
\includegraphics[width=\linewidth]{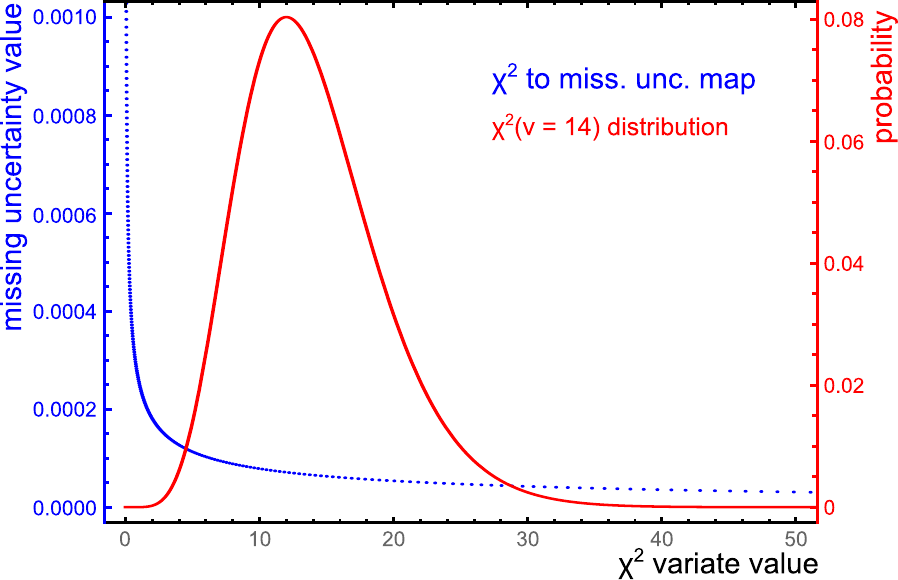}
\caption{The mapping of the $\chi^2(\nu=14)$ value to the $\sigma_{CV}$ value for our data set. By sampling the $\chi^2$ distribution, we avoided the statistical invalidity of normalizing to $\chi^2=14$.}
\label{figure-4}
\end{figure}

Despite this fact, in ignorance of the exact mechanism, the most conservative choice is to model this variation as Gaussian noise by appealing to the central limit theorem. 
We used the $\chi^2$ of our data to estimate the size of the error coming from contamination variation, assuming the same distribution for each storage time. In particular,
\begin{equation}
    \chi^2=\sum_i\frac{(\text{data}_i-\text{model}_i)^2}{\sigma_{i,\text{stat}}^2+(\sigma_{CV}\exp[(\lambda^\text{loss}_\text{Tl}-\lambda^\text{loss}_\text{Pb})t_s])^2},
\end{equation}
summing over the $i^\text{th}$ storage time where $\sigma_{i,\text{stat}}$ is the $1\sigma$ statistical uncertainty from all other components and $\sigma_{CV}$ is the $1\sigma$ statistical uncertainty to be estimated (\textit{CV}: contamination variation). 
Note the $\exp[(\lambda^\text{loss}_\text{Tl}-\lambda^\text{loss}_\text{Pb})t_s]$ factor accounts for the evolution of the initial contamination due to the different loss rates. 
To account for the fact that the $\chi^2$ of the data is itself a statistical value, we sampled the $\chi^2(\nu=14)$ distribution for each run of the Monte Carlo error propagation to determine the value of $\sigma_{CV}$ that would be used for that Monte Carlo run. 
The map between the $\chi^2(\nu=14)$ and $\sigma_{CV}$ values is shown in Fig.~\ref{figure-4}. 
Each Monte Carlo run thus had its own unique $\sigma_{CV}$ value, which was used to add variation to each data point in the Monte Carlo error propagation. This effect contributed to 69\% of our total uncertainty.

\vspace{1ex}
\textit{Nuclear shape factors}---The nuclear shape factors, $C_K$ and $C(E_e)$, which determine the $\beta_b$ decay rate and the neutrino cross section, can be expressed in terms of the appropriate nuclear matrix elements. 
For first-forbidden (ff) transitions, these are $w$, $w'$, $x$, $x'$, $u$, $u'$, $z$, $\xi'v$, and $\xi'y$ as defined in~\cite{Millener_1982,behrens1982electron,Zhi_2013}. 

In the case of $\beta_b$ decay of fully ionized $^{205}$Tl, we have~\cite{bambynek_orbital_1977}
\begin{equation}
\begin{aligned}
 C_K &=  \bigl[\xi w' +\xi' v + \tfrac{1}{3} W^b_0 w  - \tfrac{1}{3} w \bigr]^2 \\
 &+ \bigl[ \xi (x'+u') - \xi'y + \frac{1}{3} W^b_0 x + \tfrac{1}{3} (W^b_e - W^b_{\nu_e}) u \\ 
&- \tfrac{1}{3}(x+u) \bigr]^2 + \tfrac{1}{18}(W^b_{\nu_e})^2(2x+u)^2 + \tfrac{1}{12} (W^b_{\nu_e})^2 z^2 ,
\end {aligned}
\end{equation}
where  $W^b_e = 1-I(^{205}$Pb$^{81+})/m_e c^2$, with the ionization energy of $^{205}$Pb$^{81+}$, $I(^{205}$Pb$^{81+}) = 101.336$~keV. $W^b_{\nu_e}$ is the energy of the emitted neutrino in units of the electron rest mass energy, $W^b_{\nu_e}=Q_{\beta_b}/(m_e c^2)$, and  $W^b_0 = W^b_e + W^b_{\nu_e}$. $\xi = \alpha Z \lambdabar_e/(2R)$ is a measure of the Coulomb energy of a uniform sphere of radius $R$ approximating the nuclear charge distribution, where $\lambdabar_e$ is the reduced electron Compton wavelength. 

In the case of neutrino capture on neutral $^{205}$Tl to a state $j$ in $^{205}$Pb, the shape factor is given by~\cite{Millener_1982,behrens1982electron,Zhi_2013}
\begin{equation}
\label{eq:shapenu}
 C_{j,\nu}(W^\nu_e) = k + ka~W^\nu_e + \frac{kb}{W^\nu_e} +kc~(W^\nu_e)^2,
\end{equation}
where $W^\nu_e = W^\nu_{\nu_e} + (\Delta M c^2 - E^*_j)/(m_e c^2) + 1$ is the energy of the emitted electron in units of the electron rest mass energy, with $\Delta M c^2 = -50.6(5)$~keV being the difference in atomic masses between $^{205}$Tl and $^{205}$Pb. 
$W^\nu_{\nu_e}$ is the energy of the absorbed neutrino in units of the electron rest mass energy, and $E^*_j$ is the excitation energy of the final $\mathrm{^{205}Pb}$ $j$ state. The coefficients $k$, $ka$, $kb$, and $kc$ are given by
\begin{equation}
\begin{aligned}
k = & \bigl[\zeta_0^2 +\tfrac{1}{9} w^2 \bigr] + \bigl[ \zeta_1^2 + \tfrac{1}{9}(x+u)^2 -\tfrac{4}{9}\gamma_1 \mu_1 u(x+u) \\ 
& +\tfrac{1}{18}(W^\nu_0)^2(2x+u)^2 - \tfrac{1}{18} \lambda_2(2x-u)^2 \bigr] \\
& + \tfrac{1}{12}\bigl[(W^\nu_0)^2-\lambda_2\bigr]z^2, \\
ka = & \bigl[-\tfrac{4}{3} u Y -\tfrac{1}{9}W^\nu_0(4 x^2+ 5 u^2) \bigr]- \bigl[ \tfrac{1}{6}W^\nu_0z^2 \bigr], \\
kb = & -\tfrac{2}{3}\mu_1 \gamma_1 \bigl\{-[\zeta_0 w] + [\zeta_1(x+u)] \bigr\},  \\
kc = & \bigl[\tfrac{4}{9} u^2 + \tfrac{1}{18}(2x+u)^2 +  \tfrac{1}{18}\lambda_2(2x-u)^2\bigr] \\
&+ \tfrac{1}{12}(1+\lambda_2)z^2 ,
\end{aligned}
\end{equation}
with
\begin{equation}
\begin{aligned}
V =  \xi' v + \xi w', \quad Y = \xi'y - \xi (x'+u') \\
\zeta_0 = V + \tfrac{1}{3} W^\nu_0 w, \quad \zeta_1 = Y + \tfrac{1}{3} W^\nu_0 (u-x),\\
\end{aligned}
\end{equation}
where $W^\nu_0 = W^\nu_e - W^\nu_{\nu_e} = (\Delta M c^2 - E^*_j)/(m_e c^2) +1$,
$\gamma_{1} = \sqrt{1-(\alpha Z)^2}$, and $\mu_1$ and $\lambda_2$ are Coulomb functions that account for the differences in the distortion of the electron wavefunctions with different wave numbers by the Coulomb potential of the nucleus.

For heavy nuclei like the ones considered in this work, we have $\xi \gg W_0$. 
In our case, $\xi = 16.3$ and $W_0^b = 0.86$. In addition, for neutrino energies $E_\nu \ll \xi m_e c^2 \approx 8$~MeV, which includes the full energy range of  $pp$ neutrinos, it is enough to consider the relativistic matrix elements $\xi'v$ and $\xi'y$, together with those terms that are multiplied by $\xi$ in the evaluation of the shape factors for both $\beta_b$ decay and neutrino capture.  
Under this approximation, the neutrino capture shape factor becomes energy independent and we have
\begin{equation}
    C_K \approx C_\nu \approx \bigl[\xi'v + \xi w'\bigr]^2 + \bigl[\xi'y -\xi (x'+u')\bigr]^2 .
\end{equation}

Given the fact that the solar neutrino spectrum reaches neutrino energies for which the approximation above may become invalid, we consider the full energy dependence in Eq.~\eqref{eq:shapenu} when evaluating cross sections for first-forbidden transitions. 
We include first-forbidden transition to negative parity states in $^{205}$Pb, based on shell-model calculations using the \texttt{NATHAN}
code~\cite{Caurier_2005} together with the Kuo-Herling hole-hole
interaction~\cite{Warburton_1991}. 
The matrix elements have been evaluated using the quenching factors determined in~\cite{Zhi_2013} that give a value of $C_K^{\text{theo}}=8.0(24)\times10^{-3}$ for the $^{205}$Tl ($1/2^+_1$) to $^{205}$Pb ($1/2^-_1$) transition. We have assigned a relative error of 30\% to the shell-model value corresponding to the average root mean square difference obtained when computing the shape factor of the experimentally known first-forbidden beta decays of $^{205}$Au, $^{205}$Hg, $^{206}$Hg, $^{206}$Tl, and $^{207}$Tl~\cite{Zhi_2013,mancino_notitle_2024}.
The shell-model value agrees with the experimental result $C_K= 7.6(8)\times10^{-3}$ measured in this work. 

To fully account for the experimental data when describing the neutrino absorption to the first $1/2^-$ state in $^{205}$Pb, we write the neutrino capture shape factor as
\begin{equation}
    C_{1/2^-_1,\nu} (W_e) = C_{K}^{\text{exp}} + \left[C_{1/2^-_1,\nu}^\text{theo}(W_e)-C_{K}^\text{theo}\right],
\end{equation}
where the second term is evaluated using the theoretical values for the matrix elements. Based on the approximation discussed above, we expect this contribution to give a very small energy dependence to the shape factor at low energies.  
We express the energy dependent shape factor as a function of the neutrino-energy as a Taylor expansion around the mean energy of the captured $pp$ neutrinos of $E_0 = 284$ keV as
\begin{equation}
\begin{aligned}
C_{1/2^-_1}(E_{\nu_e}) =\ &  7.6(8)\times10^{-3} + 0.14(4)\times 10^{-3}\\
 & + 0.56(17) \times 10^{-3} \left(\frac{E_{\nu_e} - E_0}{1\ \text{MeV}}\right)\\ 
 & + 0.027(8)\times 10^{-3}\left(\frac{E_{\nu_e} - E_0}{1\ \text{MeV}}\right)^2.
 \end{aligned}
\end{equation}
The above expression reproduces the energy dependent shape factor for all energies relevant to the solar-neutrino spectrum. 

In addition to the first-forbidden contribution, we also consider allowed Gamow-Teller (GT) transitions based on the charge-exchange data from the $^{205}$Tl(\textit{p,n})$^{205}$Pb reaction~\cite{Krofcheck_1987}. In this case, the shape factor is energy independent and is related to the Gamow-Teller matrix element as
\begin{equation}
    C_{\text{GT},\nu} = B(GT).
\end{equation}

The cross section for neutrino capture to a specific state, $j$, in
$^{205}$Pb is given by
\begin{equation}
\sigma_{j}(E_{\nu_e}) = \frac{2 \pi^2  \hbar^3\mathrm{ln(2)}}{\mathcal{K} m_e^5 c^7}  p_e E_e F(Z,E_e) C_j(E_{\nu_e}),
\end{equation}
%
where $E_e = E_{\nu_e} - \Delta M c^2  - E^*_j + m_e c^2$ is the total energy of the emitted electron, $\Delta M c^2 = 50.6(5)$~keV is the atomic mass difference, $E^*_j$ is the excitation energy of the final $^{205}$Pb $j$ state, $p_e$ is the momentum of the emitted electron, $F(Z,E_e)$ is the Fermi function,
which we evaluate following the analytical approximation from~\cite{Schenter.Vogel:1983}, and 
$\mathcal{K} = 2\overline{\mathcal{F}t} = 6144.5 (37)$~s is the
corrected $\beta$-decay constant that includes the independent
radiative correction $\Delta^V_R$ as detailed
in~\cite{hardy_superallowed_2020}.

\begin{table}
  \caption{Contributions of individual neutrino fluxes to the solar neutrino capture rate on $^{205}$Tl expressed in SNU. These rates have to be multiplied by the neutrino oscillation survival factors. In parenthesis, the contribution to the neutrino capture of the transition to $^{205}$Pb$(1/2^-_1)$ are given.\label{table1}}
  \begin{ruledtabular}
    \renewcommand{\arraystretch}{1.2}
\begin{tabular}{ccc}
Flux & Capture rate & Survival factor\\
 \hline
${pp}$ & $113  \pm 12$ $\, (110 \pm 12)$ & $0.57 \pm 0.09$ \\
${^7\text{Be}}$ & $26 \pm 3$ $\, (23 \pm 2)$ & $0.53 \pm 0.05$ \\
${^{13}\text{N}}$  & $1.6  \pm 0.2$ $\, (1.4 \pm 0.1)$ & $0.53 \pm 0.05$ \\
${^{15}\text{O}}$  & $1.8  \pm 0.2 $\, $(1.5 \pm 0.2)$ & $0.53 \pm 0.05$ \\
${^{17}\text{F}}$  & $0.039  \pm 0.004$ $\, (0.034  \pm 0.004)$ & $0.53 \pm 0.05$ \\
${pep}$  & $1.6  \pm 0.2$ $\, (1.3 \pm 0.1)$ & $0.43 \pm 0.11$ \\
${^8\text{B}_{\text{ff}}}$  & $1.2 \pm 0.2 \, (0.55 \pm 0.07)$ & $0.32 \pm 0.02$ \\
${^8\text{B}}_{\text{GT}}$  & $34 \thickspace ^{+7} _{-5}$ $\, $ & $0.32 \pm 0.02$ \\
${hep}_{\text{ff}}$  & $0.016 \pm 0.003$ $\, (0.0058 \pm 0.0009)$ & $0.32 \pm 0.02$ \\
${hep}_{\text{GT}}$  & $0.8 \thickspace ^{+0.2} _{-0.1}$ & $0.32 \pm 0.02$ \\
\hline
${\text{Total}}$ & $180 \pm 17$ $\, (138 \pm 15)$ & - \\
\end{tabular}
\end{ruledtabular}
\end{table}

\vspace{1ex}
\textit{Individual neutrino rates}---To obtain the neutrino capture rate, we integrate over the neutrino flux $\Phi_i(E_{\nu_e})$ accounting for each individual contribution $i$ to the flux.  
Then the neutrino capture rate $R_i$  in solar neutrino units (SNUs) for a given neutrino flux
is given by
\begin{equation}
R_i =  10^{36} \times  \sum_j \int \sigma_{j}(E_{\nu_e}) \Phi_i(E_{\nu_e}) dE_{\nu_e}  \quad [\text{SNU}].
\end{equation}
We use solar
neutrino fluxes from the global analysis of solar neutrino data
including the final results of the three phases of BOREXINO (Eq.~(3.3)
of~\cite{Gonzalez-Garcia.Maltoni.ea:2024})\footnote{We have corrected a typo in the $^{17}$F flux provided in~\cite{Gonzalez-Garcia.Maltoni.ea:2024}. It should be  $5.51^{+0.75}_{-0.63}\times 10^6$~cm$^{-2}$~s$^{-1}$ instead of $5.51^{+0.75}_{-0.63}\times 10^7$~cm$^{-2}$~s$^{-1}$.}. 
We adopted the neutrino spectrum shape available on J. N. Bahcall’s website~\cite{Bahcall_site} for all the solar-neutrino fluxes, except for $^{8}$B, for which we use the accurate determination by Longfellow \textit{et al.}~\cite{Longfellow_2023}.

To account for the oscillation of the solar neutrinos, we adopted the solar neutrino survival probability factors extracted from Fig.~14.3 of~\cite{Navas_2024}, as provided by the Particle Data Group Collaboration.
We show in Table~\ref{table1} the neutrino survival factors together with the contributions of individual neutrino fluxes to the solar neutrino capture rate on $^{205}$Tl expressed in SNU \emph{without} considering neutrino oscillations. 
Furthermore, the contributions of the $^8$B and $hep$ neutrino fluxes are split into two parts. 
On the one hand, we have the contribution mediated by first-forbidden operators to states of $^{205}$Pb with I$^\pi =1/2^-,3/2^-$, and $5/2^-$.
On the other hand, as discussed above, we have evaluated the contribution mediated by the Gamow-Teller operator to states of $^{205}$Pb with I$^\pi =1/2^+$ and $3/2^+$ based on data from the $^{205}$Tl(\textit{p,n})$^{205}$Pb reaction~\cite{Krofcheck_1987}. As there are no $1/2^+$ or $3/2^+$ states known below 2.75 MeV and as the measured strength for these low energies is compatible with zero, we only consider the possibility of neutrino capture mediated by GT to $^{205}$Pb states with an excitation energy $>2.75$ MeV. 

\vfill\eject
%